**Unsupervised classification of cell imaging data using the quantization error in a Self-Organizing Map**

Birgitta Dresp-Langley*, John M. Wandeto+

*Centre National de la Recherche Scientifique CNRS UMR 7357 Université de Strasbourg, FRANCE ;
+Department of Information Technology, Dedan Kimathi University of Technology, Nyeri, KENYA

**Abstract**

This study exploits previously demonstrated properties (i.e. sensitivity to spatial extent and intensity of local image contrasts) of the quantization error in the output of a Self-Organizing Map (SOM-QE). Here, the SOM-QE is applied to double-color-staining based cell viability data in 96 image simulations. The results from this study show that, as expected, SOM-QE consistently and in only a few seconds detects fine regular spatial increase in relative amounts of RED or GREEN pixel staining across the test images, reflecting small, systematic increase or decrease in the percentage of theoretical cell viability below a critical threshold. While such small changes may carry clinical significance, they are almost impossible to detect by human vision. Moreover, here we demonstrate an expected sensitivity of the SOM-QE to differences in the relative physical luminance (Y) of the colors, which translates into a RED-GREEN color selectivity. Across differences in relative luminance, the SOM-QE exhibits consistently greater sensitivity to the smallest spatial increase in RED image pixels compared with smallest increases of the same spatial magnitude in GREEN image pixels. Further selective color contrast studies on simulations of biological imaging data will allow generating increasingly larger benchmark datasets and, ultimately, unravel the full potential of fast, economic, and unprecedentedly precise predictive imaging data analysis based on SOM-QE.

<u>Keywords</u>: Self-Organizing Maps; Quantization Error; Color Selectivity; Scanning Electron Microscopy Images; Cell Viability Imaging; Unsupervised Classification; Computation Time

**Introduction**

The quantization error in a fixed-size Self-Organizing Map (SOM) with unsupervised winner-take-all learning [1a,b] has previously been used successfully to detect meaningful changes across series of medical, satellite, and random dot images [2,3,4,5,6,7,8]. The computational properties of the quantization error in SOM are capable of reliably discriminating between the finest differences in local pixel color intensities in complex images including scanning electron micrographs of cell surfaces [9]. Moreover, the quantization error in the SOM (SOM-QE) reliably signals changes in contrast or color when contrast information is removed from, or added to, arbitrarily to images, not when the local spatial position of contrast elements in the pattern changes. While non-learning-based and fully supervised image analysis in terms of the RGB Mean reflects coarser changes in image color or contrast well enough by comparison, the SOM-QE was shown to outperform the RGB mean, or image mean, by a factor of ten in the detection of single-pixel changes in images with up to five million pixels [7,8]. The sensitivity of the QE to the finest change in magnitude of contrast or color at the single pixel level is statistically significant, as shown in our previous work [7,8]. This reflects a finely tuned color sensitivity of a self-organizing computational system akin to functional characteristics of a specific class of retinal ganglion cells identified in biological visual systems of primates and cats [10]. Moreover, the QE's computational sensitivity and single-pixel change detection performance surpasses the capacity limits of human visual detection, as also shown in our previous work [2-9].

The above mentioned properties of the SOM-QE make it a promising tool for fast, automatic (unsupervised) classification of biological imaging data as a function of structural and/or ultra-structural changes that are not detectable by human vision. This was previously shown in our preliminary work [9] on the example of Scanning Electron Micrographs (SEM) of HIV-1 infected CD4 T-cells with varying extent of virion budding on the cell surface [11,12]. SEM image technology is used in virology to better resolve the small punctuated ultra-structural surface signals correlated with surface-localized single viral particles, so-called virions [11,12]. A defining property of a retrovirus such as the HIV-1 is its ability to assemble into particles that leave producer cells, and spread infection to susceptible cells and hosts, such as CD4 lymphocytes, also termed T cells or "helper cells". This leads to the morphogenesis of the viral particles, or virions, in three stages: assembly, wherein the virion is created and essential components are packaged within the target cell; budding, wherein the virion crosses the plasma membrane (Figure 1), and finally maturation, wherein the virion changes structure and becomes infectious [11,12].

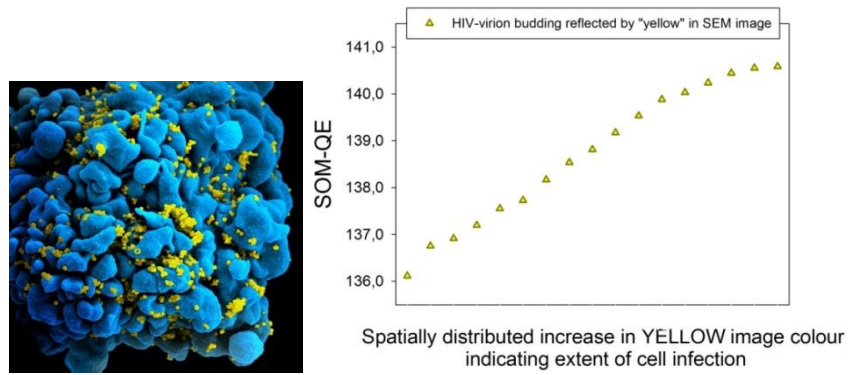

**Figure1**: Color coded SEM image of a CD4 T-cell with ultra-structural surface signals (left), in yellow here, correlated with surface-localized single HIV-1 viral particles (virions). Some of our previous work [7] had shown that SOM-QE permits the fast automatic (unsupervised) classification of sets of SEM images as a function of ultra-structural signal changes that are invisible to the human eye (right).

Another potential exploitation of SOM-QE in biological image analysis is cell viability imaging by RED, GREEN, or RED-GREEN color staining (Figure 2). The common techniques applied for determination of in vitro cell size, morphology, growth, or cell viability involve human manual work, which is imprecise and frequently subject to variability caused by the analyst himself or herself [13]. In addition, considering the necessity for evaluation of a large amount of material and data, fast and reliable image analysis tools are desirable. The use of accessible precision software for the automatic (unsupervised) determination of cell viability on the basis of color staining images would allow accurate classification with additional advantages relative to speed, objectivity, quantification, and reproducibility.

In this study here, we used SOM-QE for the fast and fully automatic (unsupervised) classification of biological imaging data in 126 simulation images. Examples of the original images used for the SOM-QE analyses here are available online at:

https://www.researchgate.net/publication/340529157_CellSurvivalDeathTrend-ColorStainingImageSimulations-2020

The test images variable RED-GREEN color staining indicative of different degrees of cell viability. For this study here, we chose variations between 44% and 56% of theoretical cell viability, i.e. variations below the threshold level that may carry clinical significance, but are not easily detected by human vision [13].

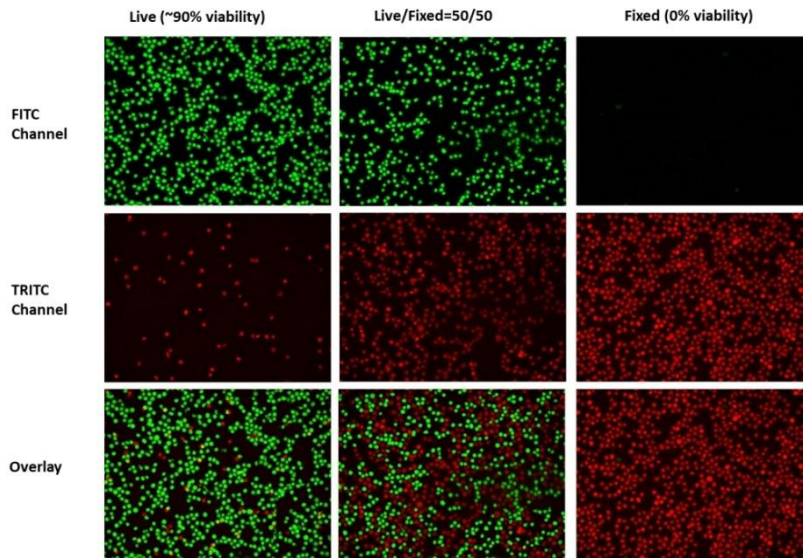

**Figure 2**: Color coded cell viability image data produced by GREEN (top), RED (middle), and RED-GREEN (bottom) staining indicating 90% (left), 50% (middle), and 0% (right) cell viability. For this study here, we generated image data using relative variability of RED and GREEN pixel color reflecting >50% and <50% variability in cell viability, and submitted the images to automatic classification by SOM-QE [2,7,8].

**Materials and Methods**

96 cell viability images with variable RED-GREEN color staining indicative of different degrees of cell viability between 50% and 56 %, indicated by an increase in the relative number of GREEN image pixels, and 44% and 50%, indicated by an increase in the relative number of RED image pixels were computer generated. All 96 images were of identical size (831x594). After training the SOM on one image (any image from a set may be used for training), the others from the set were submitted to SOM analysis to determine the SOM-QE variability as a function of the selectively manipulated image color contents, indicative of variable theoretical cell viability, expressed in percent (%).

*Images*

A cell image indicating 50% cell viability (cf. Figure 2), considered as the theoretical *ground truth* image here, displays an equivalent number, or spatial extent, of RED and GREEN dots with a specific, fixed intensity range in terms of their RGB values (here R>100<256 and G>100<256). In half of the test images from this study, the GREEN pixel contents were selectively augmented by a constant number of 5 pixels per image, yielding 48 image simulations of colour staining data indicative of a theoretical increase in cell viability from 50% to about 56%. In the other 48 images, the green pixel contents were selectively augmented by a constant number of 5 pixels per image, yielding image simulations of colour staining data indicative of a theoretical decrease in cell viability from 50% to about 44%. For a visual comparison between images reflecting 50% and 90% cell viability, based on relative amounts of combined RED and GREEN staining, see Figure 2 (bottom). Image dimensions, RGB coordinates of the selective 5-pixel-bunch RGB spatial color increments, and their relative luminance values (Y), are summarized here below in Table 1.

**Table 1:** Color parameters of the test images

| COLOR | RGBmin | RGBmax | N pixels in ground truth image/total N image pixels | N pixels • per test image | Cumulated N pixels • across test images | R G B of pixels added | Relative Luminance of pixels added Y=0.2126R+0.7152G+0.0722B |
|---|---|---|---|---|---|---|---|
| **RED** | 100, 0, 0 | 255, 0, 0 | 164 538/493 614 | +5 | +120 | 255, 0, 0 | 55.13 |
| | | | | | | 255, 65, 65 | 105.39 |
| **GREEN** | 0, 100, 0 | 0, 255, 0 | 164 538/493 614 | +5 | +120 | 0, 65, 0 | 45.79 |
| | | | | | | 65, 255, 65 | 200.89 |
| **BLACK** Background | 0, 0, 0 | 0, 0, 0 | 164 538/493 614 | 0 | 0 | -- | -- |

*SOM prototype and quantization error (QE)*

The Self-Organizing Map (a prototype is graphically represented here in Figure 3, for illustration) may be described formally as a nonlinear, ordered, smooth mapping of high-dimensional input data onto the elements of a regular, low-dimensional array [1]. Assume that the set of input variables is definable as a real vector $x$, of n-dimension. With each element in the SOM array, we associate a parametric real vector $m_i$, of n-dimension. $m_i$ is called a model, hence the SOM array is composed of models. Assuming a general distance measure between $x$ and $m_i$ denoted by $d(x,m_i)$, the map of an input vector $x$ on the SOM array is defined as the array element $m_c$ that matches best (smallest $d(x,m_i)$) with $x$. During the learning process, the input vector $x$ is compared with all the $m_i$ in order to identify its $m_c$. The Euclidean distances $\|x\text{-}m_i\|$ define $m_c$. Models that are topographically close in the map up to a certain geometric distance, denoted by $h_{ci}$, will activate each other to learn something from the same input $x$. This will result in a local relaxation or smoothing effect on the models in this neighborhood, which in continued learning leads to global ordering. SOM learning is represented by the equation

$$m(t+1) = m_i(t) + \alpha(t)h_{ci}(t)[x(t) - m_i(t)] \qquad (1)$$

where $t = 1,2,3...$is an integer, the discrete-time coordinate, $h_{ci}(t)$ is the neighborhood function, a smoothing kernel defined over the map points which converges towards zero with time, $\alpha(t)$ is the learning rate, which also converges towards with time and affects the amount of learning in each model. At the end of the *winner-take-all* learning process in the SOM, each image input vector $x$ becomes associated to its best matching model on the map $m_c$. The difference between $x$ and $m_c$, $\|x\text{-}m_c\|$, is a measure of how close the final SOM value is to the original input value and is reflected by the quantization error QE. The QE of $x$ is given by

$$QE = 1/N \sum_{i=1}^{N} \left\| X_i - m_{c_i} \right\| \tag{2}$$

where N is the number of input vectors $x$ in the image. The final weights of the SOM are defined by a three dimensional output vector space representing each R, G, and B channel. The magnitude as well as the direction of change in any of these from one image to another is reliably reflected by changes in the QE.

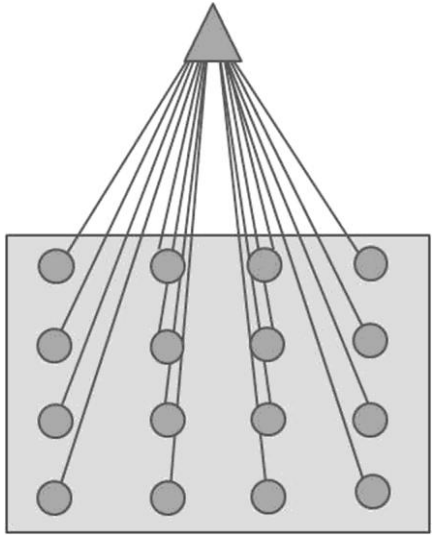

**Figure 3**: Representation of the SOM prototype with 16 models, indicated by the filled circles in the grey box. Each of these models is compared to the SOM input in the training (unsupervised winner-take-all learning) process. Here in this study, the input vector corresponds to the RGB image pixel space. The model in the map best matching the SOM input will be a winner, and the parameters of the winning model will change towards further approaching the input. Parameters of models within close neighborhood of the winning model will also change, but to a lesser extent compared with those of the winner. At the end of the training, each input space will be associated with a model within the map. The difference be-tween input vector and final winning model determines the quantization error (QE) in the SOM output.

*SOM training and data analysis*

The SOM training process consisted of 1 000 iterations. The SOM was a two-dimensional rectangular map of 4 by 4 nodes, hence capable of creating 16 models of observation from the data. The spatial locations, or coordinates, of each of the 16 models or domains, placed at different locations on the map, exhibit characteristics that make each one different from all the others. When a new input signal is presented to the map, the models compete and the winner will be the model whose features most closely resemble those of the input signal. The input signal will thus be classified or grouped in one of models. Each model or domain acts like a separate decoder for the same input, i.e. independently interprets the information carried by a new input. The input is represented as a mathematical vector of the same format as that of the models in the map. Therefore, it is the presence or absence of an active response at a specific map location and not so much the exact input-output signal transformation or magnitude of the response that provides the interpretation of the input. To obtain the initial values for the map size, a trial-and-error process was implemented. It was found that map sizes larger than 4 by 4 produced observations where some models ended up empty, which meant that these models did not attract any input by the end of the training. It was therefore concluded that 16 models were sufficient to represent all the fine structures in the image data. The values of the neighborhood distance and the learning rate were set at 1.2 and 0.2 respectively. These values were obtained through the trial-and-error method after testing the quality of the first guess, which is directly determined by the value of the resulting quantization error ; the lower this value, the better the first guess. It is worthwhile pointing out that the models were initialized by randomly picking vectors from the training image, called the "original image" herein. This allows the SOM to work on the original data without any prior assumptions  about a level of organization within the data. This, however, requires to start with a wider neighborhood function and a bigger learning-rate factor than in procedures where initial values for model vectors are pre-selected [1 b]. The procedure described here is economical in terms of computation times, which constitutes one of its major advantages for rapid change/no change detection on the basis of even larger sets of image data before further human intervention or decision making. The computation time of SOM analysis of each of the 98 test images to generate the QE distributions was about 12 seconds per image.

**Results**

After SOM training on the reference image (unsupervised learning), the system computes SOM-QE for all the images of a given series in a few seconds, and writes the SOM-QE obtained for each image into a data file. Further steps generate output plots of SOM-QE, where each output value is associated with the corresponding input image. The data are plotted in increasing/decreasing orders of SOM-QE magnitude as a function of their corresponding image variations. Results are shown here below for the two test image series (Figure 4). The SOM-QE is plotted as a function of increments in the relative number, by adding pixel bunches of constant size and relative luminance, of GREEN or RED image pixels. For the corresponding image parameters and variations, see Table 1.

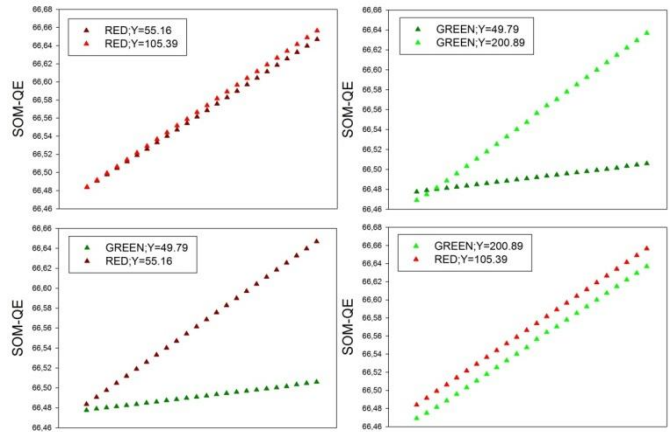

**Figure 4**: SOM-QE classification of the 96 test images, with the SOM-QE plotted as function of increasing or decreasing theoretical cell viability indicated by a small and regular increase in the spatial extent of green or red pixels across the corresponding images. The data show the expected SOM-QE sensitivity to the relative luminance (Y) of a given color (top), and its color selectivity (bottom). For any relative luminance Y, the color RED, by comparison with the color GREEN, is signaled by QE distributions of greater magnitude. Future studies on a wider range of color-based imaging data will allow to further benchmark SOM-QE color selectivity.

**Conclusions**

In this work we exploit the RED-GREEN color selectivity of SOM-QE [7,8] to show that the metric can be used for a fast, unsupervised classification of cell imaging data where color is used to visualize the progression or remission of a disease or infection on the one hand, or variations in cell viability before and after treatment on the other. Similarly successful simulations were obtained previously on SEM images translating varying extents of HIV-1 virion budding on the host cell surface, coded by the color YELLOW, in contrast with healthy surface tissue, coded by the color BLUE [9]. Our current work, in progress, reveals hitherto un-suspected properties of self-organized [14] mapping, where the potential of the SOM_QE is revealed in terms of a computational tool for detecting the finest clinically relevant local changes in large series [15] of imaging data. Future image simulations will allow further benchmarking of SOM-QE selectivity for increasingly wider ranges of color variations in image simulations of biological data. This should, ultimately, result in providing a basis for the automatic analysis of biological imaging data where information relative to contrast and/or color is exploited selectively to highlight disease-specific changes in organs, tissue structures, or cells.